\documentclass[review]{elsarticle}

\usepackage{lineno,hyperref}
\usepackage{multirow}
\usepackage{algorithm}
\usepackage{algorithmic}
\usepackage{amsmath,amssymb,amsfonts}
\usepackage{graphicx}
\usepackage{float}
\usepackage{textcomp}
\usepackage{xcolor}
\usepackage{color}
\usepackage{subfigure}
\usepackage{bm}
\usepackage{makecell} 
\usepackage{diagbox} 

\newcommand\algorithmictuning{\textbf{Tuning params:}}
\newcommand\TUNING{\item[\algorithmictuning]}

\journal{Journal of Information Processing and Management}

\begin{document}

\begin{frontmatter}

\title{Eating Healthier: Exploring Nutrition Information for Healthier Recipe Recommendation}


\author[address1]{Meng Chen}
\ead{mchen@sdu.edu.cn}

\author[address1]{Xiaoyi Jia}
\ead{jiaxiaoyishd@163.com}

\author[address2]{Elizabeth Gorbonos}
\ead{gorb6620@mylaurier.ca}

\author[address2]{Chnh T. Hong}
\ead{choang@wlu.ca}

\author[address3]{Xiaohui Yu}
\ead{xhyu@yorku.ca}

\author[address2]{Yang Liu\corref{mycorrespondingauthor}}
\ead{yangliu@wlu.ca}
\cortext[mycorrespondingauthor]{This is to indicate the corresponding authors.}

\address[address1]{Shandong Univeristy, 27 Shanda South Road, Jinan, Shandong, China}
\address[address2]{Wilfrid Laurier University, 75 University Avenue West, Waterloo, Ontario, Canada}
\address[address3]{York University, 4700 Keele Street, Toronto, Ontario, Canada}

\begin{abstract}
With the booming of personalized recipe sharing networks (e.g., Yummly), a deluge of recipes from different cuisines could be obtained easily. In this paper, we aim to solve a problem which many home-cooks encounter when searching for recipes online. Namely, finding recipes which best fit a handy set of ingredients while at the same time follow healthy eating guidelines. This task is especially difficult since the lions share of online recipes have been shown to be unhealthy. In this paper we propose a novel framework named NutRec, which models the interactions between ingredients and their proportions within recipes for the purpose of offering healthy recommendation. Specifically, NutRec consists of three main components: 1) using an embedding-based ingredient predictor to predict the relevant ingredients with user-defined initial ingredients, 2) predicting the amounts of the relevant ingredients with a multi-layer perceptron-based network, 3) creating a healthy pseudo-recipe with a list of ingredients and their amounts according to the nutritional information and recommending the top similar recipes with the pseudo-recipe. We conduct the experiments on two recipe datasets, including Allrecipes with 36,429 recipes and Yummly with 89,413 recipes, respectively. The empirical results support the framework's intuition and showcase its ability to retrieve healthier recipes.
\end{abstract}


\end{frontmatter}


\section{Introduction}\label{sec:introduction}
The presence of food related content on the web has become prominent in recent years. Along with the overwhelming volume of images on social-media, the use of online sites as a source for recipes and culinary ideas is expanding \cite{KissingCuisines, min2018you, freyne2011recipe}. With the large quantity of available data, finding the right recipe becomes a difficult task. Food recommender systems, a domain specific subclass of recommender systems, have been designed to assist in providing users with recipes or meal plans that accommodate their preferences. Food recommenders, which usually rely on a tally to indicate people's aggregated opinion; however are inevitably facing the following challenges that cannot be easily addressed by adopting existing methods. First, the majority of online recipes do not adhere to the nutritional guidelines set by international food agencies and are therefore regarded as unhealthy. Second, a negative correlation is displayed between rating scores and healthiness measures, meaning highly rated recipes tend to be even less healthy \cite{Trattner2017}. Third, multiple aspects influence our food preferences among which are geographical and cultural affiliation. Being linked to these factors, cuisine recommendation generated based on people's votes can be a great mean to express group inclinations; however, the level of healthy and individual preference should be captured in other forms.

In contrast, a user's culinary inclination can be perceived as a result of personal ingredient preferences or some latent factors. Some works use ingredients' likings, whether in the form of an explicit input \cite{RecipeCompletion} or by indirectly learning them from recipe ratings \cite{Freyne2010}; while others employ standard collaborative filtering approaches to uncover hidden factors \cite{Trattner2017}. Despite their merits in recommending recipes one would like or suggesting suitable extensions to an ingredient list, they do not consider the \textit{recipe-completion} task (where a complete recipe is constructed from a list of ingredients) and for the most part they overlook the nutritional aspect.

To meet these challenges, we propose NutRec, a framework that tackles the completion task whilst striving to retrieve healthy recipes. While in previous studies the relationships between ingredients are investigated and result in ingredient recommendations for the purpose of list-completion or substitution \cite{RecipeCompletion, IngredientNetworks}, here we aim to generate full healthy recipe recommendations. Specifically, we first propose an embedding-based ingredient predictor to predict the relevant ingredients with some given ingredients, and then utilize a multi-layer perceptron-based predictor to capture the quantities of supplemented ingredients. Using an iterative process, with nutritional compatibility considered at each step, we generate a healthy pseudo-recipe with a set of ingredients and the corresponding amounts. This pseudo-recipe is then matched against the recipes in the dataset to find similar healthy recommendations. The datasets used in this paper are composed of real-world online recipes collected from two major food websites, namely Allrecipes.com \cite{Allrecipes} and Yummly.com \cite{Yummly}. The contributions of this paper are summarized as follows.
\begin{itemize}
\item We propose a healthy recipe recommendation framework (NutRec), which first builds a healthy pseudo-recipe considering the nutritional values and then scans the recipe dataset for items resembling the pseudo-recipe. Our proposed NutRec relies not only on the relations between the ingredients themselves but also on those of their quantities, which ultimately dictate the healthiness of a recipe. To the best of our knowledge, no prior study has incorporated these features.
\item The pseudo-recipe is a list of ingredients with their quantities, and the nutritional values of the pseudo-recipe should match the predefined targets as best as possible. To generate the pseudo-recipe, we first propose an embedding-based ingredient predictor, which embeds all the ingredients into a latent space and predicts the supplemented ingredients based on the distances of ingredient representations; we then propose an amount predictor to compute the quantities of the supplemented ingredients.
\item We conduct extensive experiments with two real recipe datasets, and the experimental results confirm the superiority of our methods over the baselines. To facilitate the community research, we have publicly released the datasets \footnote{https://github.com/mchenSDU/NutRec}.
\end{itemize}

The rest of the paper is structured as follows. Section \ref{sec:related-work} presents prior research done in the field, and Section \ref{sec:nutrec} introduces our proposed framework NutRec. In Section \ref{sec:evaluation} we describe the evaluation methods and discuss the attained results, and finally in Section \ref{sec:conclusion} we conclude the findings and propose future endeavors.

\section{Related Work}\label{sec:related-work}
Recently, recommendation has been a hot research problem with the booming of social networks. Scholars have carried out many research works on different fields \cite{cui2017augmented, chen2018mpe,chen2014nlpmm,cui2018distribution,chen2016jcst}, e.g., location recommendation, movie recommendation. In this paper, we focus on recipe recommendation and discuss the recent progress of the related methods and applications.

\subsection{Recipe recommendation}\label{sec:related-work-recipe-recommendation}
Recipe recommendation is a sub-domain of the larger food recommendation task, and the problem of recommending healthy recipes has been tackled in numerous studies. In literature, \textit{healthiness} is measured by comparing macro-nutrient values against several international guidelines. The World Health Organization (WHO) and the United Kingdom Food Service Agency (FSA) guidelines are adopted in \cite{Trattner2017,Elsweiler2017,Howard2012}. Furthermore, the Nutrient Reference Values of Australia and New Zealand (NRV) is incorporated in \cite{Elsweiler2015}. Within the realm of healthy recipe recommenders, the datasets explored vary between pre-existing healthy recipes (\cite{Freyne2010, fudholi2009ontology, espin2016nutrition, tumnark2013ontology}) and online recipe datasets (\cite{Trattner2017,Elsweiler2017,Elsweiler2015}). Online data repositories pose a greater challenge as they hold an inherent tendency for unhealthiness. In what follows, we summarize the related recommendation methods according to the data types.

\subsubsection{Recommendation with pre-existing healthy recipes}
Traditionally, recommendation methods rely on ratings and strive to suggest recipes the user would rate high. For example, Freyne and Berkovsky \cite{Freyne2010} use recipe ratings to explore the relationship between healthy recipes and their ingredients. They devise a break-down technique and examine the feasibility of extracting the users' ingredient preferences from their ratings on recipes. They demonstrate their method can surpass a collaborative filtering baseline approach. However, when dealing with online recipes, users' ratings on recipes are not good indicators for healthiness, as many recipes with high ratings are pretty unhealthy.

Furthermore, some methods rely on domain experts' advice and are generally knowledge-based. Fudholi et al. \cite{fudholi2009ontology} design an ontology-based daily menu assistance system to meet daily energy requirements. Espin et al. \cite{espin2016nutrition} present a nutritional recommender system, helping elderly users to draw up their own healthy diet plans following the nutritional experts' guidelines. Tumnark et al. \cite{tumnark2013ontology} describe a food and nutrition ontology working with a rule-based knowledge framework to satisfy the specific dietary needs of weightlifters. These studies use predefined diets and menus which are all considered appropriate but not necessarily compatible with the user's needs. The recommender component in those systems may accept, reject or adjust the menus based on rules set by nutrition experts or clinical guidelines. Unfortunately, the advantages of this methodology cannot be effortlessly applied to online recipes. Online data are not guarantied to include all required ontology properties nor are all recipes valid candidates (i.e. being healthy).

\subsubsection{Recommendation with online recipe datasets}
To improve the healthiness of the recommendations, multiple methods are presented with online recipe datasets. For instance, Trattner and Elsweiler \cite{Trattner2017} apply post-filtering on the generated recommendations and re-rank them to make the healthier recipes appear on the top of the prediction list. Seeing that online repositories often contain multiple versions of the same dish, Elsweiler and Morgan \cite{Elsweiler2015} propose a solution which diverges from the quest to find a single healthy recipe and instead aims to generate daily meal plans. These plans are balanced recipe combinations which are obtained via exhaustive search. Further, they \cite{Elsweiler2017} evaluate the possibility of substituting one recipe with a similar yet healthier alternative. In addition, Trattner et al. \cite{tratter2019} analyze the upload behavior of an online food community platform, and find that social connections in the form of online friendship relations bear a great potential to predict what types of recipes user are going to create in the future and which ingredients are going to be used. The research contributes to a better understanding in online food behavior and benefits healthy recipe recommendation.

In this paper, we focus on online recipes and try to tackle the challenge from another angle. Our method first generates a healthy recipe-like draft and then searches the dataset for matchings. Unlike other papers, it does not rely on ratings nor on the availability of nutritional information for the recipes.

\subsection{Recipe completion}\label{sec:related-work-recipe-completion}
Recipe completion requires an understanding of the relations between ingredients in a recipe. A recent extensive survey preforms such an investigation with the aim of characterizing different cuisines \cite{KissingCuisines}. Kikuchi et al. \cite{Kikuchi2017} demonstrate the effects of seasonality on recipes' content and illustrate how ingredient co-appearance in recipes changes throughout a year. In \cite{Yokoi2015} the typicality of ingredients within a recipe category (e.g. ``hamburger'') is defined. This measure is used to help determine how well a specific ingredient fits into a target recipe category.

The two most relevant works to this paper are \cite{RecipeCompletion} and \cite{IngredientNetworks}. The \textit{complementary network}, introduced in \cite{IngredientNetworks}, models all of the recipes' ingredients as vertices of a graph. The weights between any ingredient pair represent the probability of co-appearance. To extract insights from this model they use clustering algorithms on the network. Even though they set out to solve different problems, e.g. substituting ingredients in recipes, this network model lends itself nicely to a completion task. Out of all the related papers, \cite{RecipeCompletion} is the only one to intently deal with the completion task. The tools they employ are matrix factorization (\textit{MF-based}) models, which have two major components including non-negative matrix factorization and two-step least root squares. These methods are able to extract compatibility information from the recipe-ingredient matrix.

\section{Proposed Framework: NutRec}\label{sec:nutrec}
\subsection{General framework}
\begin{table}[!t]
\centering
\caption{{ {Notations and descriptions.}}}
\renewcommand{\arraystretch}{1.0}
\begin{tabular}{ l| l}
\Xhline{1pt}
Notations & Descriptions \\
\hline
$r, i$ &  recipe, ingredient  \\
$N_{r}$ &  the number of ingredients in recipe $r$\\
$\bm{v}_{i}$  &  the embedding vector of target ingredient $i$\\
$\bm{c}_{i}$  &  the contextual vector of target ingredient $i$\\
$\bm{v}'_{i}$  &  the embedding vector of contextual ingredient $i$\\
$d$ & the dimension of latent space\\
\hline
$N$ &  the set of macro-nutrients\\
$t$ & the macro-nutrients' target values\\
$p$ & the predicted macro-nutrients' values\\
\hline
$n$ & the number of ingredients adding to the pseudo-recipe\\
$k$ & the number of recommended healthy recipes\\
\Xhline{1pt}
\end{tabular}
\label{tab:notation}
\end{table}	
The proposed framework NutRec is inspired by a human-like deliberation and seeks to generate healthy recommendations over an online non-healthy dataset. The user's preferences are accommodated in the form of an ingredient-list and are passed to the framework as input. NutRec consists of three main stages: 1) predicting the relevant ingredients with an ingredient-list; 2) predicting the amounts of the relevant ingredients; 3) creating a healthy pseudo-recipe and searching for similar recipes in the dataset. The working process of NutRec is presented in full in Algorithm \ref{alg:recommend}. { {Tab.~\ref{tab:notation} lists the notations used in this paper.}}
\begin{itemize}
\item \textbf{Ingredient predictor}. We propose an embedding-based predictor to find candidate relevant ingredients by computing top compatible ingredient lists with the current ingredient-set $prSet$ (line 3).
\item \textbf{Amount predictor}. A list of ingredients does not suffice to calculate nutritional values, to accomplish this task we demand the rations. We develop an amount predictor which enables us to predict the quantities of a given ingredient-set (line 4).
\item \textbf{Healthy recipe recommendation}. By using the ingredient predictor and the amount predictor, the \textit{FindBestIngredientToAdd} assess how each candidate ingredient impacts the nutritional values and returns the best ingredient to append to the working set. We iteratively execute this process for $n$ times, and generate a healthy pseudo-recipe, i.e. a data-structure containing the ingredients along with their respective amounts (line 9). Finally, with the healthy pseudo-recipe, we search the dataset for the top-$k$ most similar recipes (line 10).
\end{itemize}

\begin{algorithm}[!t]
 \caption{NutRec: Recommend}\label{alg:recommend}
 \begin{algorithmic}[1]
 \renewcommand{\algorithmicrequire}{\textbf{Input:}}
 \renewcommand{\algorithmicensure}{\textbf{Output:}}
 \REQUIRE 	$s$: the initial ingredient-set\\
 			$k$: { {the number of recommended healthy recipes}}\\	
 \ENSURE $l$: the list of recommended recipes\\
 \TUNING $n$: { {the number of ingredients adding to the pseudo-recipe}}\\
		$cos$: cosine-weight for similar recipes search\\
  \STATE $prSet = s$
  \FOR {$i = 0$ to $n$}
    \STATE $candidateIngredients$ = IngredientPredictor($prSet$)
    \STATE $amountofCandidateIngredients$ = AmountPredictor($prSet$)
	\STATE $bestIngredient$ = FindBestIngredientToAdd($candidateIngredients$,
	\STATE $amountofCandidateIngredients$)
    \STATE $prSet$ += $bestIngredient$
  \ENDFOR
  \STATE $pseudoRecipe$ = CreateRecipeFromSet($prSet$)
  \STATE $l$ = FindSimilarRecipes($pseudoRecipe, cos, k$)
 \RETURN $l$
 \end{algorithmic}
 \end{algorithm}

\subsection{The ingredient predictor}

\subsubsection{model description}
In order to retrieve the most befitting ingredients, we propose an embedding-based ingredient predictor (IP-embedding), in which the ingredients are projected into a latent space and the ones that usually occur in a recipe are close to each other. Inspired by the recent progress of deep learning and neural networks \cite{fernandez2018prospect,mikolov2013distributed,bagheri2018neural}, we propose to use a distributed representation method to model the generation of the given recipe. Given a recipe $r: \{i_1, i_2, \cdots, i_{N_r}\}$ containing $N_r$ ingredients, the objective function is to maximize the probability of each target ingredient $i_a$ given its corresponding context information:
\begin{equation}
\label{onerecipe}
g(i_a) =  Pr(i_a|Context({i_a})),
\end{equation}
where we consider the ingredients except $i_a$ in the recipe $r$ as its context.

Formally, we model each target ingredient $i_a$ with a $d$-dimensional embedding vector $\bm{v}_{i_a} \in \mathbb{R}^{d}$ and embed each contextual ingredient into the same latent vector space. Given an ingredient $i_a$, we average the embedding vectors of multiple contextual ingredients to compute a contextual vector $\bm{c}_{i_a}$,
\begin{equation}
\label{contextVec}
\bm{c}_{i_a} =  \dfrac{1}{N_r - 1} \sum_{t=1, t \neq a}^{N_r} \bm{v}'_{i_t},
\end{equation}
where $\bm{v}'_{i_t}$ is the embedding vector of the contextual ingredient $i_t$. Note that, we distinguish between the role of a target ingredient and a contextual ingredient, and represent the same ingredient $i_a$ using two different vectors ($\bm{v}_{i_a}$ and $\bm{v}'_{i_a}$) depending on which role it takes. We then apply a softmax function to compute the probability of a target ingredient $i_a$ given its contextual vector as follows:
\begin{equation}
\label{softmax}
Pr(i_a|Context({i_a})) = \dfrac{\exp(\bm{c}_{i_a}^{\mathrm{T}} \cdot \bm{v}_{i_a})} {\sum_{i \in {\cal I}} \exp(\bm{c}_{i_a}^{\mathrm{T}} \cdot \bm{v}_{i})},
\end{equation}
where ${\cal I}$ is the set of ingredients. Such a model connects the target ingredient and its contexts via the embedding representations.

For parameter learning, we need to maximize the log probability defined in Eq.~(\ref{onerecipe}) over all the recipes. However, directly optimizing this objective function is impractical because the cost of computing the full softmax for the multiclassifier to predict the target ingredient is extremely high. Therefore, we adopt the efficient and effective negative sampling strategy \cite{mnih2012fast,Levy2014Neural} to approximate the full softmax. When we train the vector of $i_a$ , we first obtain a corresponding negative sample set $NEG(i_a)$, in which $i_x$ is not the same as $i_a$ if $i_x \in NEG(i_a)$. Then we define
\begin{equation}
L^{i_a}(i) =
\begin{cases}
1, \quad i = i_{a}; \\
0, \quad i \neq i_{a},
\end{cases}
\end{equation}
where $L^{i_a}(i)$ is the label of the ingredient $i$. The labels of positive instances are 1, and 0 otherwise.

Given the contextual ingredients of a target ingredient $i_a$, we want to maximize the occurrence probability of $i_a$ and meanwhile minimize the occurrence probability of negative samples
$i_x \in NEG(i_a)$. The objective function for the target ingredient $i_a$ therefore becomes
\begin{equation}
g(i_a) = \prod_{i \in {i_a} \bigcup NEG(i_a)} Pr(i|Context({i_a})),
\end{equation}
where
\begin{equation}
Pr(i|Context({i_a})) =
\begin{cases}
\begin{aligned}
\sigma(\bm{c}_{i_a}^{\mathrm{T}} \bm{v}_{i}), &\quad L^{i_a}(i) = 1; \\
1-\sigma(\bm{c}_{i_a}^{\mathrm{T}} \bm{v}_{i}), &\quad L^{i_a}(i) = 0,
\end{aligned}
\end{cases}
\end{equation}
in which $\sigma(z)= {(1+\exp(-z))}^{-1}$ is the sigmoid function.

Further,
\begin{equation}
\label{loss1}
\begin{aligned}
g(i_a) &= \prod_{i \in {i_a} \bigcup NEG(i_a)} Pr(i|Context({i_a})),\\
&= \prod_{i \in {i_a} \bigcup NEG(i_a)} [\sigma(\bm{c}_{i_a}^{\mathrm{T}} \bm{v}_{i})]^{L^{i_a}(i)} \cdot [1-\sigma(\bm{c}_{i_a}^{\mathrm{T}} \bm{v}_{i})]^{1-L^{i_a}(i)},\\
&=\sigma(\bm{c}_{i_a}^{\mathrm{T}} \bm{v}_{i_a}) \prod_{i \in NEG(i_a)} (1-\sigma(\bm{c}_{i_a}^{\mathrm{T}} \bm{v}_{i})).\\
\end{aligned}
\end{equation}

Finally, we define the object function for all the ingredients,
\begin{equation}
\begin{aligned}
\cal L &= \sum_{r \in {\cal R}} \sum_{a=1}^{N_r} \log g(i_a),\\
 &= \sum_{r \in {\cal R}} \sum_{a=1}^{N_r} { \left( \log[\sigma(\bm{c}_{i_a}^{\mathrm{T}} \bm{v}_{i_a})] + \sum_{i \in NEG(i_a)} \log[1-\sigma(\bm{c}_{i_a}^{\mathrm{T}} \bm{v}_{i})] \right)},\\
 &= \sum_{r \in {\cal R}} \sum_{a=1}^{N_r} { \left( \log[\sigma(\bm{c}_{i_a}^{\mathrm{T}} \bm{v}_{i_a})] + \sum_{i \in NEG(i_a)} \log[\sigma(-\bm{c}_{i_a}^{\mathrm{T}} \bm{v}_{i})] \right)}.\\
\end{aligned}
\end{equation}

Given an incomplete recipe, we regard all its ingredients as the context, and compute $Pr(i|Context(i))$ based on Eq.~(\ref{softmax}) for each candidate ingredient $i$, and return the top-$N_i$ ingredients as outputs.

\subsubsection{Parameter learning}
IP-embedding needs to learn the embedding vectors of all the ingredients. We represent all the parameters with $\Theta$, and learn the IP-embedding model by using maximum a posterior (MAP):
\begin{equation}
\Theta = \arg \max \sum_{r \in {\cal R}} \sum_{a=1}^{N_r} \log g(i_a)  - \dfrac{1}{2} \lambda||\Theta||^{2} ,\\
\end{equation}
where $\lambda||\Theta||^{2}$ is the regularization term.

All parameters are trained using the stochastic gradient ascent method. During the training process, the algorithm iterates over the target ingredients of all the recipes. At each time, a target ingredient $i_a$ with its contexts is used for update. For a certain context $Context(i_a)$, the target ingredient $i_a$ is a positive sample, and we randomly sample $N_e$ unobserved ingredients as negative samples. After computing the loss function, the error gradient is obtained via backpropagation and we use the gradient to update the parameters in our model.
\begin{equation}
\begin{aligned}
\bm{v}_{i_a} &= \bm{v}_{i_a} + \eta \left( [1- \sigma(\bm{c}_{i_a}^{\mathrm{T}} \bm{v}_{i_a})] \cdot \bm{c}_{i_a} - \lambda \bm{v}_{i_a} \right), \\
\bm{v}_{i_x} &= \bm{v}_{i_x} + \eta \left(-\sigma(\bm{c}_{i_a}^{\mathrm{T}} \bm{v}_{i_x}) \cdot \bm{c}_{i_a} - \lambda \bm{v}_{i_x} \right), i_x \in NEG(i_a),\\
\bm{v}'_{i_c} &= \bm{v}'_{i_c} + \eta ( [1- \sigma(\bm{c}_{i_a}^{\mathrm{T}} \bm{v}_{i_a})] \cdot \bm{v}_{i_a} - \sum_{i_x \in NEG(i_a)} [\sigma(\bm{c}_{i_a}^{\mathrm{T}} \bm{v}_{i_x}) \cdot \bm{v}_{i_x}] - \lambda \bm{v}'_{i_c} ),\\
 &\qquad i_c \in Context(i_a),\\
\end{aligned}
\end{equation}
where $\eta$ is the learning rate.

\subsection{The amount predictor}\label{sec:amount predictor}
This component is tasked with predicting the gram-amounts a set of ingredients are likely to be used in a recipe. When evaluating the nutritional values of an ingredient-set, we need to know the proportions among the ingredients. Since a programmatic approach to this problem is likely to result in non-realistic values (e.g., a kilo of onions and 10 grams of flour), to gain intuition about real-world ratios we have built an amount predictor.

The amount predictor is based on a dense one hidden layer network with ReLU activation function. The input and output layers' dimensions equate to the number of unique ingredients in the dataset ($m$), and the optimal size for the hidden layer is determined by tuning. The input for the network is a binary vector $\vec{x}$ of length $m$ with the value 1 assigned to the ingredients which are included in the set and 0 to the rest. The output of the network is a numerical vector $\vec{x'}$ of the same size with the amount values (in grams) for all ingredients.

The amount predictor is trained using the Keras framework\footnote{F. Chollet, 2015. Keras. https://keras.io}. A binary recipe-ingredient matrix $X$ is utilized as the input along with an additional numerical matrix $X'$ of recipe-ingredient-amounts which serves as the target.

\subsection{Healthy recipe recommendation}
\subsubsection{Finding the best supplemented ingredients and generating the pseudo-recipe}
In order to effectively capture the user's preference and maintain the healthiness, we evaluate the content similarity between the target nutrients and each candidate set. In particular, Mean Squared Errors (\textit{MSE}) measure is adopted as the distance measure. The error in this scenario is the distance between the target macro-nutrient values and the predicted ones. Formally:
\begin{equation}
MSE = \frac{1}{|N|} \sum_{i \in N}(t_i - p_i)^2,
\label{eq:mse}
\end{equation}
\noindent where $N$ is the set of macro-nutrients (protein, carbohydrates, sugars, fat, saturated fat, sodium and fiber), $t$ is the macro-nutrients target values and $p$ is the predicted macro-nutrients values of the current set extended by the inspected candidate. The ingredient which results in the lowest MSE score is chosen.\\

The process halts after $n$ iterations or when no new ingredient can improve upon the error. In this work we have limited the number of iteration to 5, so as to not exceed the average number of recipe ingredients (as we will show in Tab.~\ref{tab:datasets}). Additionally, we do not wish to encourage the algorithm to pick ingredients which typically appear in small amounts (and thus do not meaningfully influence the MSE) just for the sake of adding more ingredients. The final product of this step is our pseudo-recipe.

\subsubsection{Retrieving similar recipes}
We point out our aspiration is to find recipes which resemble the pseudo-recipe both in the ingredients themselves but also their quantities. Unfortunately, these two properties do not necessarily coincide. To mitigate this issue, we prescribe \textit{similarity} as a weighted average of the Jaccard and cosine metrics:
\begin{equation}
\text{sim} = \text{COS}_\text{weight} \cdot \text{sim}_\text{cos} + (1-\text{COS}_\text{weight}) \cdot \text{sim}_\text{jaccard}
\label{eq:sim}
\end{equation}

Cosine similarity captures the likeness in terms of amounts:
\begin{equation}
\text{sim}_\text{cos} = \frac{\vec{a} \cdot \vec{b}}{|\vec{a}| \cdot |\vec{b}|} =
\frac{\sum_{j = 1}^{m}\vec{a}_j \cdot \vec{b}_j}
{\sqrt{\sum_{j = 1}^{m}\vec{a}_j^2} \sqrt{\sum_{j = 1}^{m}\vec{b}_j^2} },
\end{equation}
\noindent where $\vec{a}$ is the pseudo-recipe amounts vector and $\vec{b}$ is a recipe amounts vector, both of length $m$ (the number of unique ingredients in the dataset). Jaccard index is used to emphasize shared ingredients:
\begin{equation}
\text{sim}_\text{jaccard} = \frac{|A \cap B|}{|A \cup B|}.
\end{equation}
\noindent Here $A$ and $B$ are the ingredient-sets of the pseudo-recipe and an examined recipe, respectively.

\section{Experimental Results}\label{sec:evaluation}

\subsection{Datasets}\label{sec:datasets}
The datasets utilized in this paper are collected from Allrecipes\footnote{https://www.allrecipes.com} and Yummly\footnote{https://www.yummly.com}. All nutritional values for the recipes are available on the websites. The ingredients and amounts are parsed in a semiautomatic manner, using fuzzy matching between the recipes' ingredients and the foods in the USDA dataset\footnote{US Department of Agriculture, Agricultural Research Service, Nutrient Data Laboratory. USDA National Nutrient Database for Standard Reference, Release 28. Version Current: September 2015. Available: http://www.ars.usda.gov/ba/bhnrc/ndl}. About 80\% of ingredients are recognized and at least 70\% are parsed. A parsed ingredient is a recognized ingredient which we are able to quantify in grams. To make our models more robust, we filter those recipes that only contain one ingredient. We have publicly released our datasets to facilitate the community research \footnote{https://github.com/mchenSDU/NutRec}.

\begin{figure}
  \centering
  \includegraphics[width=0.5\textwidth]{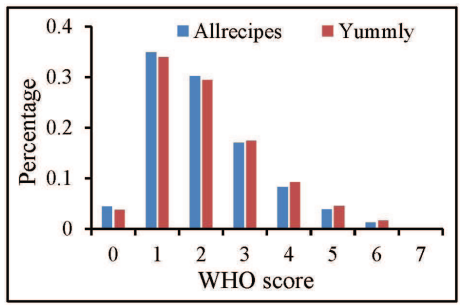}
  \caption{WHO score distribution in Allrecipes and Yummly datasets. WHO score reflects how well a recipe meets the WHO guideline on the scale of 0-7 (poorly-well). A recipe with a high WHO score is considered healthy.}
  \label{fig:datasets-who}
\end{figure}

Tab.~\ref{tab:datasets} displays the characteristics of the datasets. The WHO scores are computed using the manner described in \cite{Trattner2017}, and their distribution is displayed in Fig.~\ref{fig:datasets-who}. Since healthy recipes entail high WHO scores, it can be clearly seen that most recipes in the datasets are unhealthy. Lastly, nutrition information for ingredients is incorporated via the USDA dataset.

\begin{table}[!t]
\caption{Data statistics}
\begin{center}
\begin{tabular}{l|c|c}
\Xhline{1pt}
 & \textbf{Allrecipes}& \textbf{Yummly} \\
\hline
Number of recipes& 36,429& 89,413  \\
Average no. of ingredients per recipe & 7.2 & 7.6\\
Recognized ingredients & 82\%  & 81\%\\
Parsed ingredients & 77\% & 71\%\\
Unique ingredients & 465 & 512\\
\Xhline{1pt}
\end{tabular}
\label{tab:datasets}
\end{center}
\end{table}

For both datasets, we randomly split them into three collections in proportion of 8:1:1 as the training set, validation set, and test set, and perform 5 runs (with the same
data split) to report the average of the results. All the experiments are done on a 3.4GHz Intel Core i7 PC with 16GB main memory.

\subsection{Evaluation on the ingredient predictor}
The proposed ingredient predictor could be applied in the task of recipe completion and is examined by the following method. For each recipe, a random ingredient is removed and the ability of the models to retrieve it is tested. This method is also employed for our baseline predictors \cite{RecipeCompletion, gorbonos2018nutrec}. With IP-embedding, we first build a contextual vector $c_i$ based on Eq.~(\ref{contextVec}), and then compute $Pr(i|c_i)$ for each candidate ingredient $i \in {\cal I}$ based on the following function:
\begin{equation}
Pr(i|c_i) \propto {\bm{c}_i}^T \cdot \bm{v}_i.
\end{equation}
Finally, we predict the missing ingredient according to the probability. The results of this evaluation are reported as:
\begin{itemize}
\item Percent of predictions with the removed ingredient ranked $\leq$ 10,
\item Mean rank of the removed ingredient,
\item Median rank of the removed ingredient.
\end{itemize}
An optimal predictor would return the missing ingredient at the top of the list. Here we consider it a good prediction if the removed item appears within the first 10 results. Therefore, we would like the mean and median ranks to be low and the percent of good predictions to be as high as possible.

\subsubsection{Baselines}
We choose three methods as the baselines and compare our IP-embedding with them.
\begin{itemize}
\item \textbf{Graph-based ingredient predictor (IP-graph)} is a graph-based model $G(V,E)$, where $V$ consists of all the unique ingredients. Edges $E$ connect any two ingredients that co-appear in at least one recipe and the edge weights are the co-appearances' counts. IP-graph returns the top-$N_i$ candidate ingredients from a vertex list sorted by the combined edge weights \cite{gorbonos2018nutrec}.
\item \textbf{Multi-layer perceptron-based ingredient predictor (IP-MLP)} uses the same network structure as that of the proposed amount predictor. The output of the amount predictor is a numerical vector $\vec{x'}$ of the same size with the amount values (in grams) for all ingredients. IP-MLP returns the top-$N_i$ ingredients with the highest values in $\vec{x'}$ as the candidate ingredients.
\item \textbf{Nonnegative matrix factorization-based ingredient predictor (IP-NMF)} is inspired by the nonnegative matrix factorization model originally described in \cite{RecipeCompletion}. IP-NMF decomposes the recipe-ingredient matrix into two smaller matrices $W$ and $H$ with dimensions $u \times c$ and $c \times m$, where $c$ is the number of latent factors, $u$ is the number of recipes and $m$ is the number of unique ingredients. Further, it computes a resulting vector with $W$ and $H$, and returns the top-$N_i$ predictions correspond to ingredients with the highest values in the vector.
\end{itemize}

\begin{figure}[!t]
\centering
\subfigure[Accuracy of rank $\leq$ 10]{
\includegraphics[height=0.15\textheight]{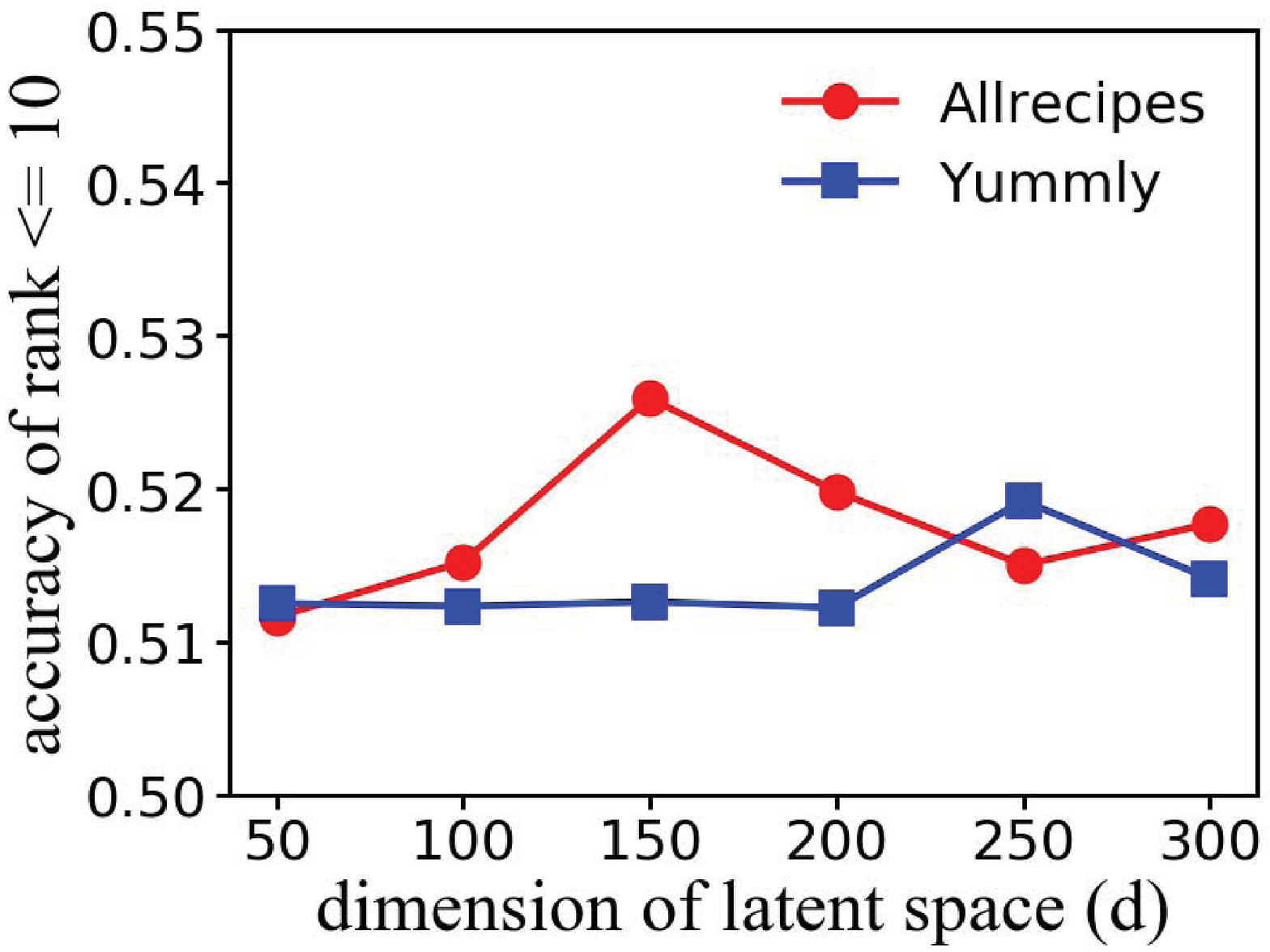}}
\subfigure[Mean rank]{
\includegraphics[height=0.15\textheight]{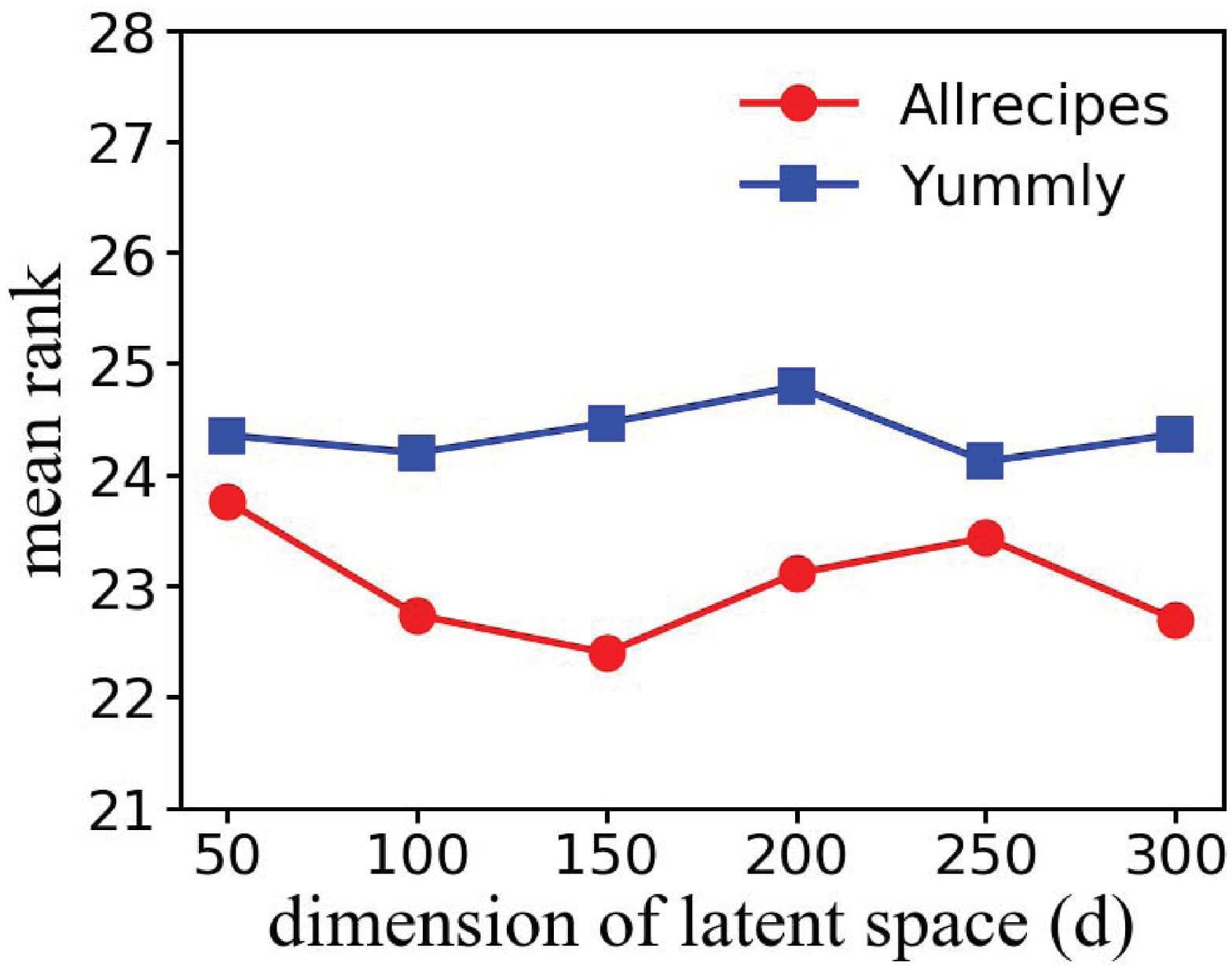}}
\subfigure[Median rank]{
\includegraphics[height=0.15\textheight]{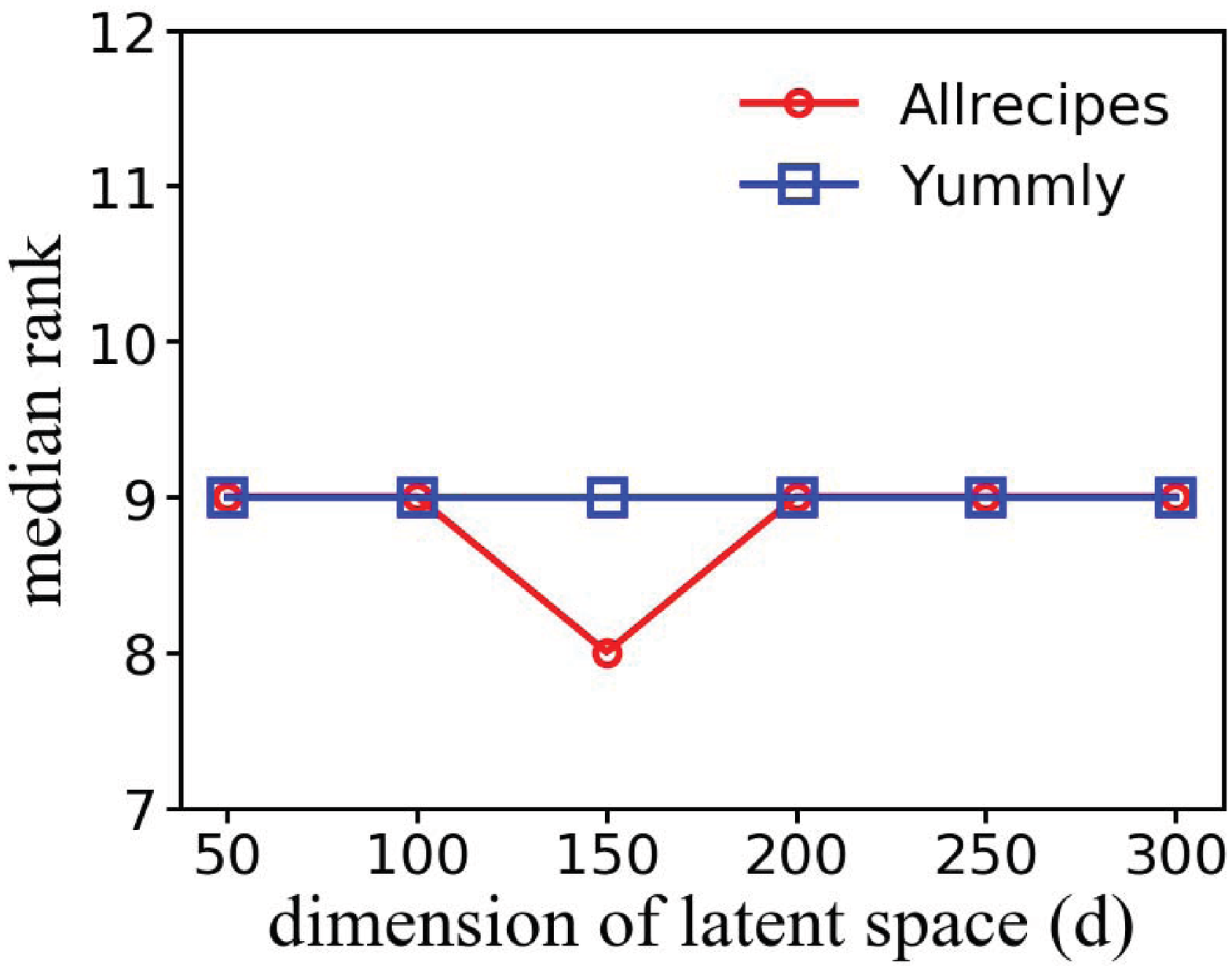}}
\caption{Empirical performance of tuning the dimension of latent space (IP-embedding). }
\label{fig:d}
\end{figure}

\subsubsection{Parameter setting and tuning}
We measure the effect of the dimension of latent space $d$ for IP-embedding. We increase $d$ from 50 to 300 with a step of 50, and demonstrate the results in Fig.~\ref{fig:d}. The performance of IP-embedding changes little with each $d$. When $d$ is equal to 150, IP-embedding performs the best on the Allrecipes dataset, and IP-embedding reaches the optimal when $d$ is 250 on the Yummly dataset. The optimal parameter is used in all subsequent experiments.

\begin{table}[!t]
\caption{Similar ingredients discovered by IP-embedding with a given ingredient}
\small
\begin{center}
\begin{tabular}{l|l|l|l}
\Xhline{1pt}
 & \textbf{pasta}& \textbf{beef}  & \textbf{vegetable oil}\\
\hline
& spaghetti& venison  & canola oil\\
 & bagels & mutton  & olive oil\\
 & cheese ravioli & lean ground beef &  lard\\
top-10 & noodles & ground turkey   &  oil \\
similar & bologna & beef sausage  & coconut oil \\
ingredients & vermicelli & turkey sausage  & large carrots\\
 & hearts of palm & roast beef    &  salted butter\\
& asparagus & rolls  & margarine\\
 & artichoke hearts & breakfast sausage   & finely chopped onion\\
& rolls & steak  &   eggplant\\
\Xhline{1pt}
\end{tabular}
\label{tab:ip4}
\end{center}
\end{table}

As a side contribution, IP-embedding could embed the ingredients into a latent space, and a pair of ingredients with more relatedness should be projected into closer embedding vectors. To validate this semantic correlation, we list the top-10 similar ingredients for a given ingredient. Here we choose ``pasta'', ``beef'', and ``vegetable oil'' as the examples of query ingredient, and return the top-10 similar ingredients based on the cosine similarity between the embedding vectors, as shown in Tab.~\ref{tab:ip4}.
\begin{itemize}
\item Given the query ingredient ``pasta'', the most similar ingredient is ``spaghetti'', which is the same food as ``pasta''. Furthermore, the top-10 similar ingredients of ``pasta'' are mainly cooked wheaten food.
\item The top-10 similar ingredients of ``beef'' are all meat or meat sausage, e.g., ``venison'', ``mutton'', and ``beef sausage''.
\item The top-5 similar ingredients of ``vegetable oil'' contain multiple kinds of oil, e.g., ``canola oil'', ``olive oil'', and ``lard''.
\end{itemize}
We can see that the similar ingredients discovered by IP-embedding demonstrate unique characteristics about the query ingredient in an explainable manner. Therefore, the learned vectors of ingredients include the semantic correlation and are pretty valuable for many tasks, e.g., ingredient clustering.

\begin{table}[!t]
\caption{Performances of the four ingredient predictors}
\small
\begin{center}
\begin{tabular}{c|c|c|c|c|c}
\Xhline{1pt}
& & IP-graph & IP-MLP & IP-NMF &  \textbf{IP-embedding}\\
\hline
\multirow{3}{*}{\textbf{Allrecipes}} & Rank $\leq$ 10 & 45.5\% & 31.0\% & 49.5\% & \textbf{52.6\%} \\

 							& Mean rank & 29.6 & 44.9 & 37.2 & \textbf{22.4}\\

		 					& Median rank & 12 & 27 & 11 & \textbf{8} \\
\hline
\multirow{3}{*}{\textbf{Yummly}} & Rank $\leq$ 10 &  43.5\% & 28.0\% & 50.0\% & \textbf{51.9\%} \\

 							& Mean rank & 32.4 & 49.8 & 37.4  & \textbf{24.1}\\

		 					& Median rank & 13 & 30 & 11 & \textbf{9} \\
\Xhline{1pt}
\end{tabular}
\label{tab:models}
\end{center}
\end{table}

\subsubsection{Comparison with existing methods}
\label{fourIP}
Finally, we compare our proposed IP-embedding with these baselines. For IP-NMF, we test it with 2-4 components (i.e., the number of hidden factors). IP-NMF obtains the best performance when the number of hidden factors is 2. For IP-MLP, we set the size of hidden layer at 128, 256, 512, respectively, and find that IP-MLP performs the best when the size of hidden layer is 128. We summarize the results of the four ingredient predictors in Tab.~\ref{tab:models}. The best results are highlighted in boldface. IP-graph constructs a graph based on the recipes and it obtains the decent performance. IP-MLP builds a simple network and it predicts the missing ingredient based on the predicted amount of each candidate ingredient. IP-MLP performs the worst as it leverages the indirect information. IP-NMF predicts the missing ingredient based on non-negative matrix factorization, and it performs better than IP-graph and IP-MLP in terms of accuracy of rank $\leq$ 10. Our proposed IP-embedding produces the best results out of the four models, as it models the relation between the target ingredient and its contexts, and predicts the missing ingredient according to the context information. It is able to include the correct ingredient within the top-10 elements in 52.6\% of the cases on the Allrecipes dataset and 51.9\% of the cases on the Yummly dataset. Its mean and median scores are also lower than the competitors. Note that, the improvements of our IP-embedding over all the baselines are statistically significant in terms of paired t-test \cite{hull1993using} with p value $<$ 0.01.

\subsection{Evaluation on the amount predictor}
\begin{figure}[!t]
\centering
\subfigure[Allrecipes]{
\includegraphics[width=0.48\textwidth]{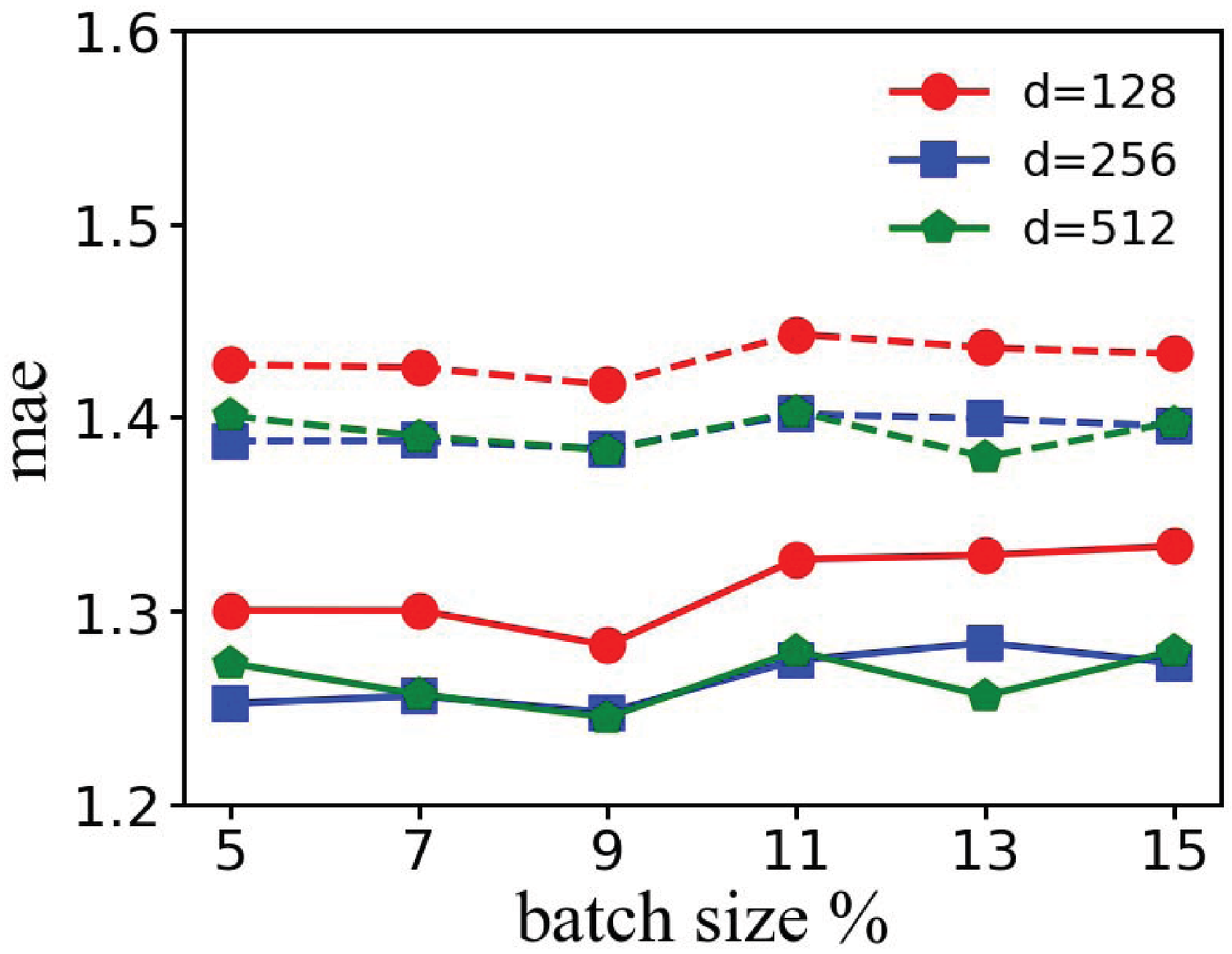}}
\subfigure[Yummly]{
\includegraphics[width=0.48\textwidth]{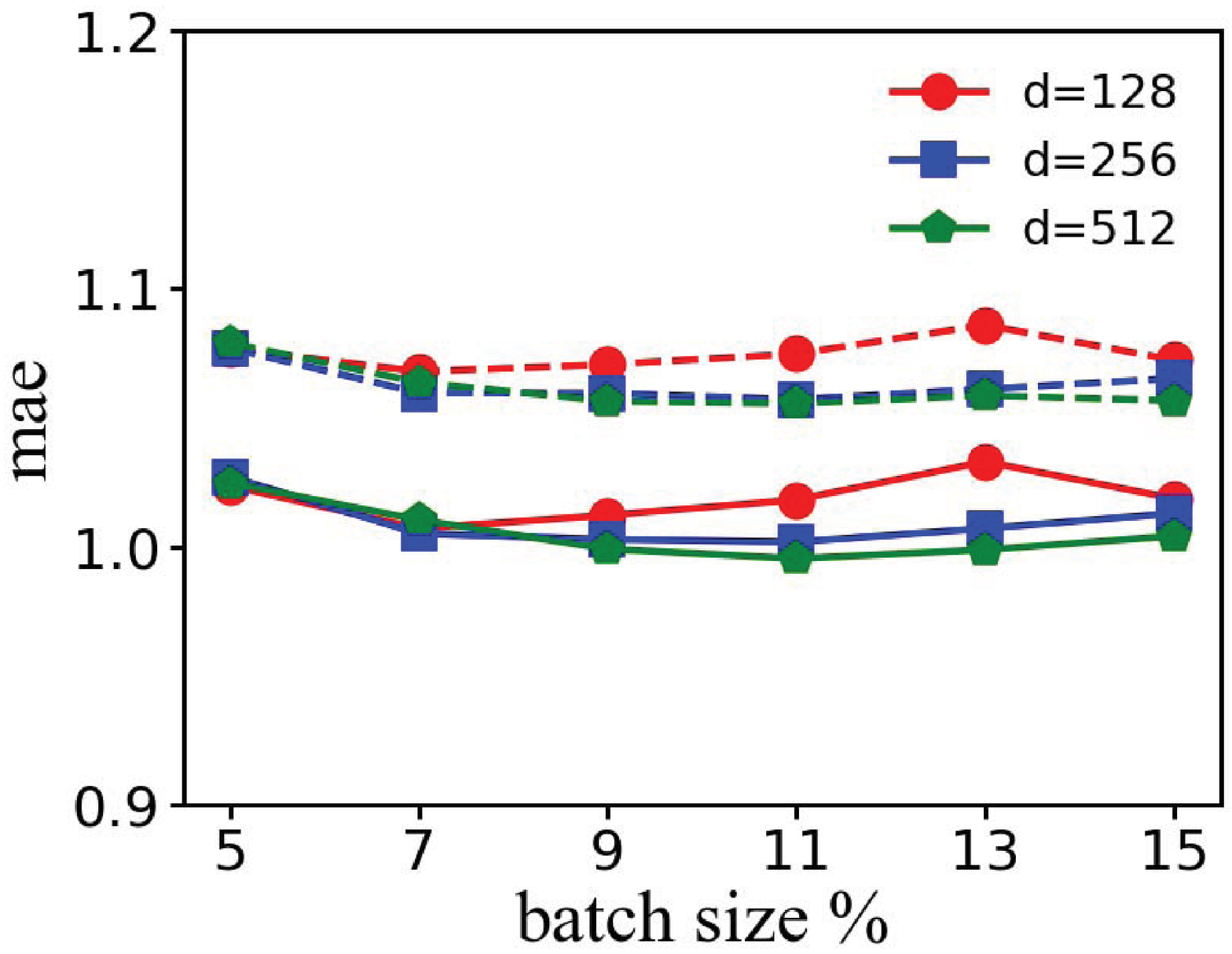}}
\caption{The performance of the amount predictor for Allrecipes and Yummly. Solid lines represent the mae on the training set and dashed lines represent the mae on the validation set.}
\label{fig:amounts-eval}
\end{figure}

The amount predictor is evaluated by the mean absolute error (mae). The evaluation is performed on the training set and the validation set while varying the size of the hidden layer and the batch size. Fig.~\ref{fig:amounts-eval} illustrates these results. A batch size of 9\% produces the best results on both datasets. The lowest train and validation mae scores for Allrecipes are 1.25 and 1.38 respectively with a hidden layer dimension of 512. For Yummly the lowest mae scores obtained are 1.0 for the training set and 1.06 on the validation set with a hidden layer the size of 512 neurons.  It is not surprising a lower validation mae is observed on Yummly as its size provides more training opportunities. 

\subsection{Evaluation on the NutRec}

As no state-of-the-art method for comparison is available, we formulate an alternative design and validate whether our proposal could recommend healthier recipes through empirical analysis. The evaluation of the framework NutRec is performed using 120 most frequent item sets. Frequent sets are sets of ingredients which repeatedly appear together in recipes. The parameters are set to $n$=5 and $k$=10. Meaning, NutRec increases the initial ingredient set by 5 ingredients and returns the top-10 most similar healthy recipes.

To measure the performance of NutRec, we adopt the mean WHO score, which is based on the recommendations' WHO score distribution. We compare our NutRec with a \textit{random-draw} baseline. The random-draw is computed using all recipes containing the initial sets. We average over these recipes' WHO scores to represent the baseline. In addition to the mean WHO score, we have studied another error-based metric, using logic similar to Eq.~(\ref{eq:mse}), in our preliminary experiments. However, our tests proved this measure to be inappropriate for precise targets. This is due to the fact that even though wide gaps between the macro-nutrient targets and the actual recipe values induce large errors, a combination of minor deviations can have the same effect.

The target macro-nutrient values are set as detailed in Tab.~\ref{tab:targets}. The table displays the seven macro-nutrients along with their recommended portion of daily energy intake. The targets are expressed in grams. Whenever a range is clearly defined we use its mean value. For upper or lower bounds (such as the case with sugar, sodium and fiber) we use a 1.5 multiplier. The target values are based on a 2000 Kcal intake and a conventional conversion is used: a single gram of protein, carbohydrates or sugar amounts to 4 Kcal and one gram of fat or saturated fat is equal to 9 Kcal.

\begin{table}[!t]
\caption{Target macro-nutrients}
\begin{center}
\begin{tabular}{c|c|c}
\Xhline{1pt}
\textbf{Nutrient} & \textbf{WHO guideline}& \textbf{Target}$^{\mathrm{a}}$ \\
\hline
Protein & 10\%-15\% & 12.5\% = 62.5g\\
Carbohydrates& 55\%-75\% & 65\% = 325g\\
Sugar& $<$ 10\%& 6.6\% = 33.3g\\
Fat& 15\%-30\%& 22.5\% = 50g\\
Saturated fat& $<$ 10\%& 6.6\% = 14.8g\\
Sodium& $<$ 2g & 1.33g\\
Fiber& $>$ 25g& 37.5g\\
\Xhline{1pt}
\multicolumn{3}{l}{$^{\mathrm{a}}$Percentage conversion is based on 2000 Kcal/day diet.}
\end{tabular}
\label{tab:targets}
\end{center}
\end{table}

\begin{figure}[!t]
\centering
\subfigure[Allrecipes]{
\includegraphics[width=0.48\textwidth]{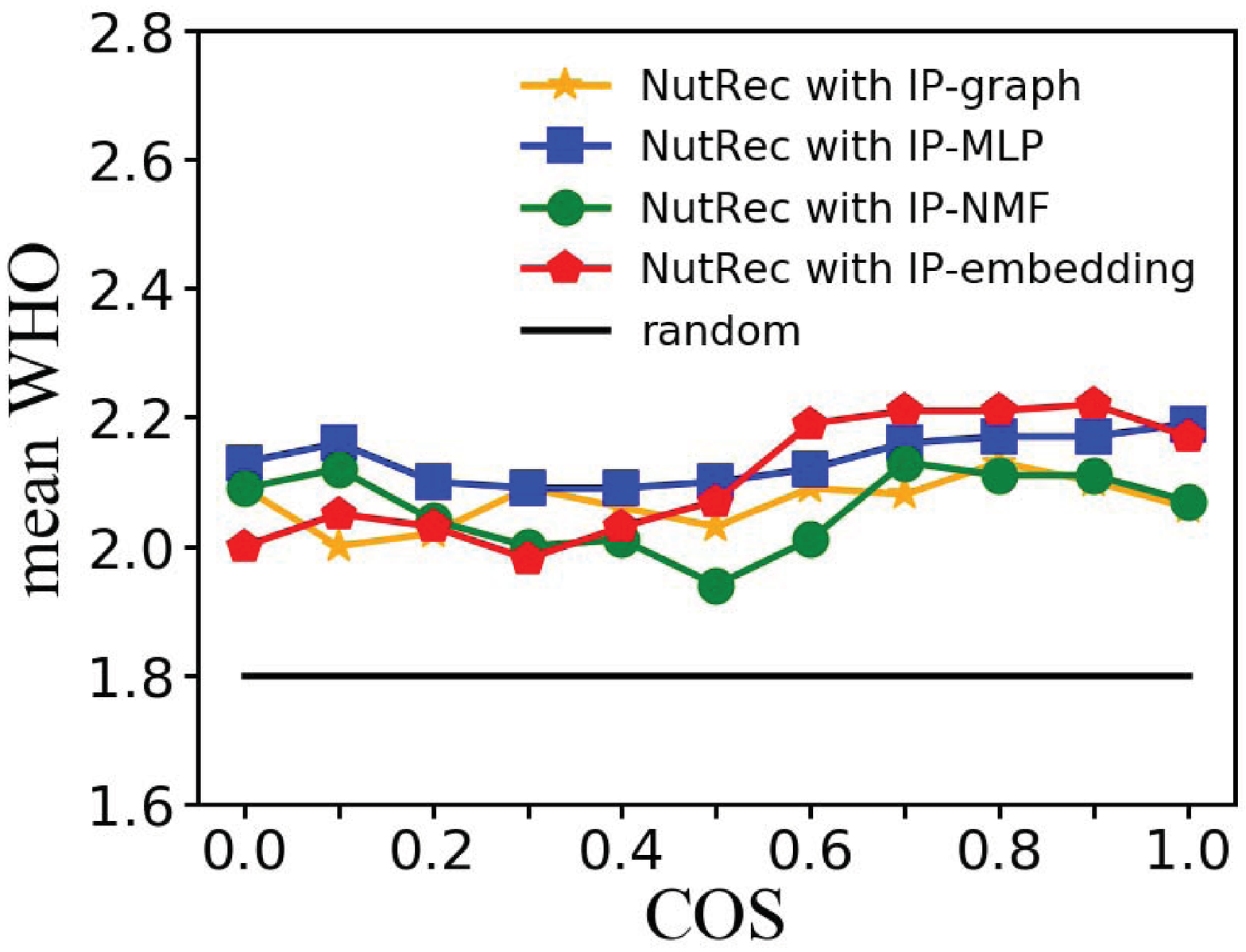}}
\subfigure[Yummly]{
\includegraphics[width=0.48\textwidth]{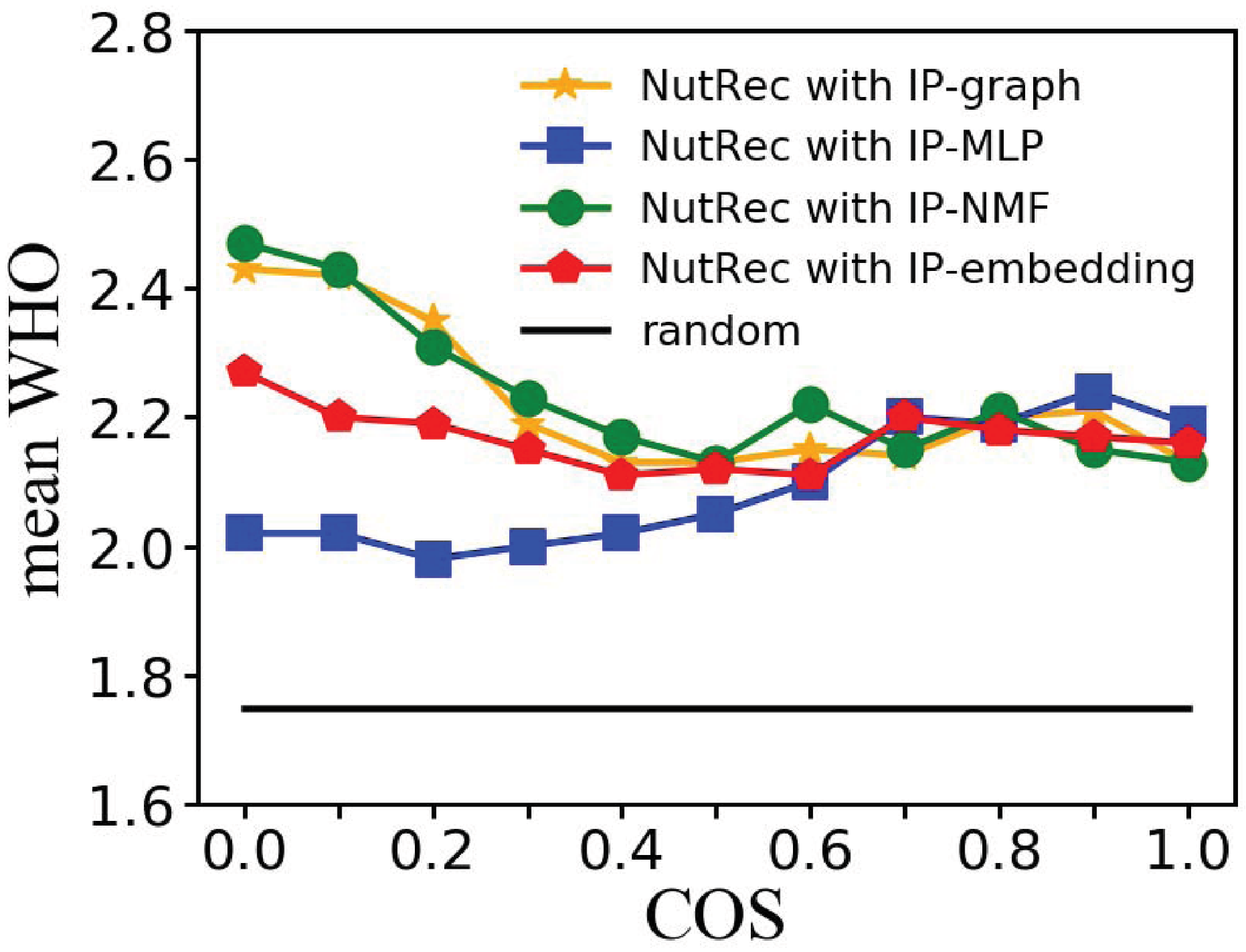}}
\caption{Mean WHO score of recommended recipes using different ingredient predictors.}
\label{fig:mean-who}
\end{figure}

We evaluate the performance of four ingredient predictors in Section~\ref{fourIP}, and we incorporate them into the framework NutRec respectively to validate the effectiveness. The mean WHO scores of the predictions as a function of $\text{COS}_\text{weight}$ (COS) are demonstrated in Fig.~\ref{fig:mean-who}. The colors correspond to the NutRec with the four ingredient predictors. IP-graph is denoted by a yellow line, IP-MLP is blue, IP-NMF is green and IP-embedding is red. The random-draw distribution is laid out by the black line. This chart illustrates that all COS settings improve the random-draw baseline, meaning our NutRec indeed recommends healthier recipes for users. For example, the baseline mean WHO scores for the top-120 most frequent sets in Allrecipes and Yummly are 1.80 and 1.75 respectively; the NutRec with IP-embedding reaches 2.22 with COS equal to 0.9 on the Allrecipes dataset and the NutRec with IP-NMF could reach 2.47 on the Yummly dataset. Additional investigation reveals that higher cosine-similarity weights actually preformed slightly better for most of the cases, meaning that the amounts of ingredients play a pivotal role in the healthy recipe recommendation. Furthermore, the NutRec with our proposed IP-embedding performs relatively better, especially on the Allrecipes dataset.

\begin{figure}[!t]
\centering
\subfigure[Allrecipes (WHO $>$ 3)]{
\includegraphics[width=0.48\textwidth]{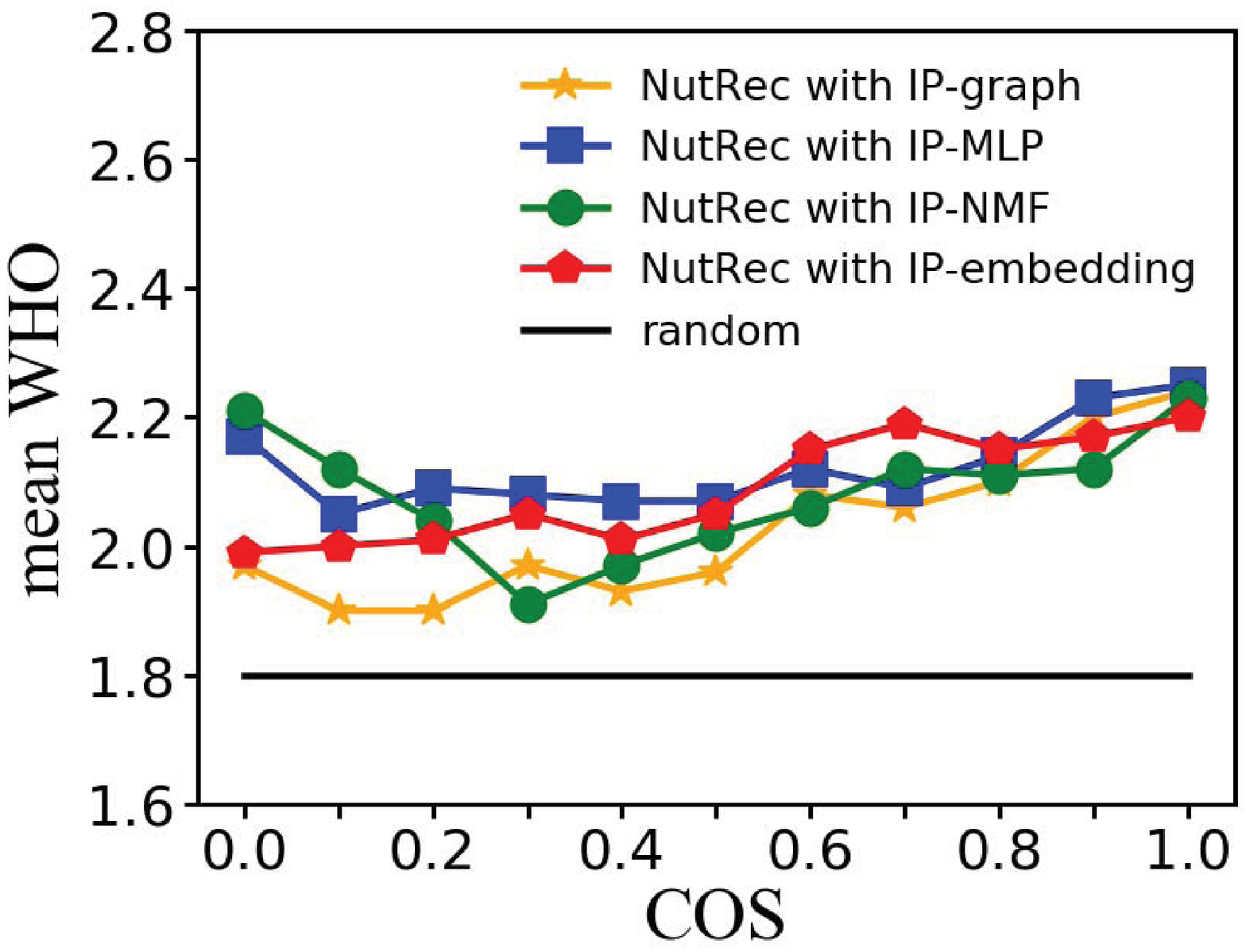}}
\subfigure[Yummly (WHO $>$ 3)]{
\includegraphics[width=0.48\textwidth]{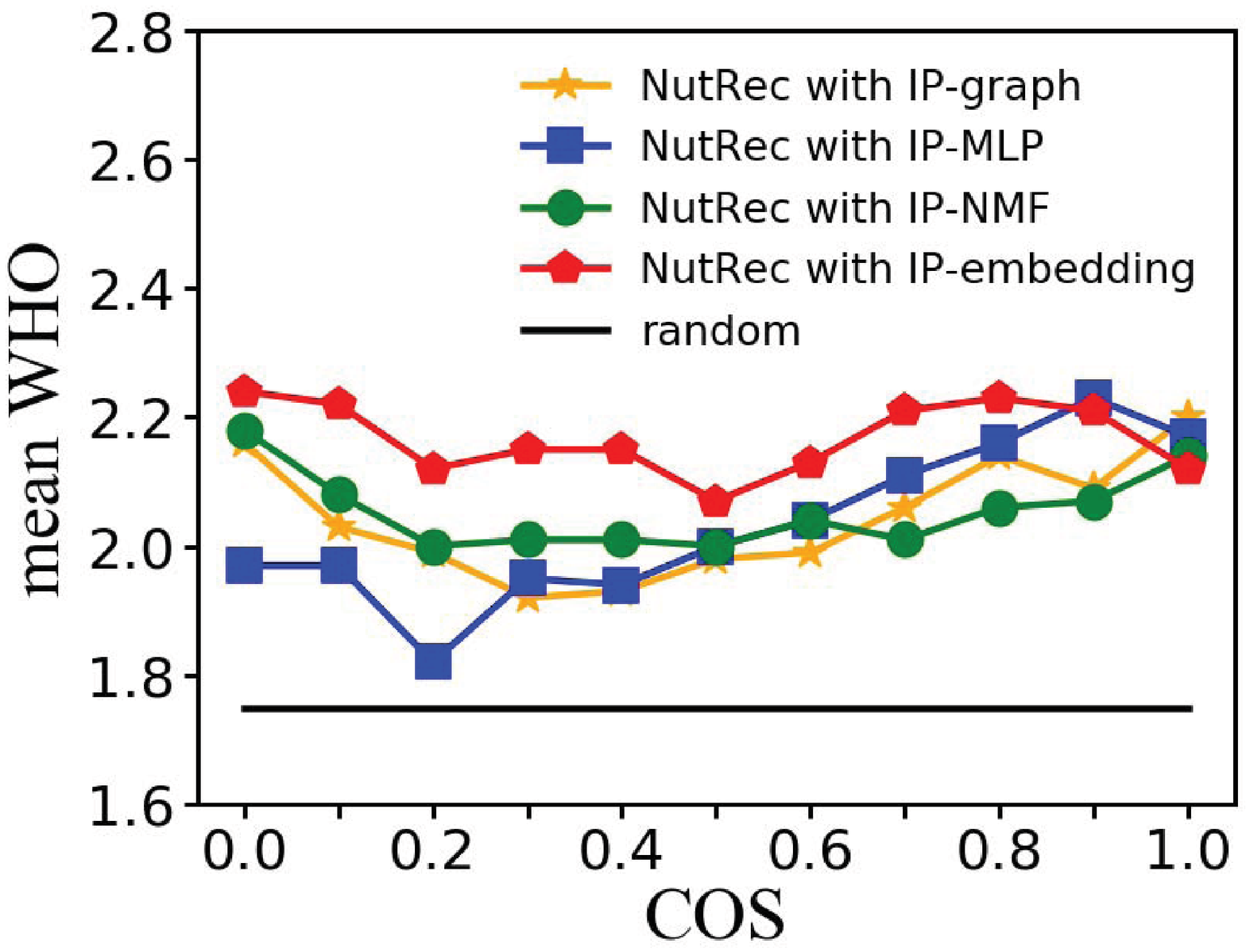}}
\caption{Mean WHO score of recommended recipes, using different ingredient predictor, when training the amounts network on recipes with WHO $>$ 3. }
\label{fig:mean-who-filtered}
\end{figure}

Further, we explore whether the mean WHO scores can be increased by training the amount predictor using only recipes with the WHO scores greater than 3 (i.e. using a dataset with mostly healthy examples). The results are illustrated in Fig.~\ref{fig:mean-who-filtered}. The actual improvement according to the mean WHO score is negligible. In most of the cases, the mean WHO scores are lower than those in Fig.~\ref{fig:mean-who}. We will point out two striking trends apparent in the results. First, as can be deduced from Fig.~\ref{fig:datasets-who}, imposing a minimum WHO score limit drastically cuts down the training data and subsequently impairs the model's predictive power. It is plausible the model's accuracy deterioration leads to incorrect ingredient combinations, for instance the algorithm may produce an ingredient-set often found in unhealthy recipes. This is evidenced in Fig.~\ref{fig:mean-who-filtered} by the inferior mean WHO scores for low COS values (that is when recommending recipes based on primarily similar ingredients). A second feature, which can be recognized in Fig.~\ref{fig:mean-who-filtered}, is the positive correlation of COS and mean WHO score, unlike the vague relation seen in Fig.~\ref{fig:mean-who}. This implies that under this setup the predicted amounts better model healthy ingredients-ratios. Notwithstanding, the fact this experiment is not able to improve the results on Yummly suggests learning using healthy recipes may be less beneficial than having a large training set.

\section{Conclusion and Future Work}\label{sec:conclusion}
In this paper, we have proposed a framework named NutRec to tackle the healthy recipe recommendation problem. NutRec first predicts the relevant ingredients and their amounts with some user-defined initial ingredients, and then creates a healthy pseudo-recipe considering the nutritional values. Finally, NutRec searches the top similar healthy recipes based on the pseudo-recipe in the dataset. As evident from the presented results, NutRec is able to improve the average healthiness of the recommended recipes (for a top-10 task) without requiring any pre-computed nutritional information for the recipes.

In the future, we would like to advance our study on the following topics. First, one technique to improve the performance may be via an ensemble model which will combine all four ingredient predictors. Second, in this paper healthy recipes are the focal point, but as the method strives to find recipes which best match a given nutrient target, it is credible it could prove useful for other types of goals, e.g., low-fat or high-protein diets. Finally, as users' ratings are a prominent part of recommender systems, an interesting extension for this work would be investigating a fashion to integrate ratings. Once such technique is applicable, it will be possible to compare the proposed algorithm with traditional recommender methods.

\bibliography{mybibfile}

\end{document}